\documentclass[preprint, raferee]{aastex62}
\usepackage{epsfig,amsmath,amsfonts,amssymb,graphicx,subfigure,enumitem,color}
\usepackage{amssymb}
\usepackage{amsmath}
\usepackage[varg]{txfonts}
\usepackage{natbib}
\usepackage{url}             
\usepackage{graphicx}
\graphicspath{{figures/}}
\usepackage{float}
\newcommand{\beq}{\begin{equation}}
\newcommand{\eeq}{\end{equation}}
\newcommand{\beqa}{\begin{eqnarray}}
\newcommand{\eeqa}{\end{eqnarray}}
\newcommand{\beqann}{\begin{eqnarray*}}
\newcommand{\eeqann}{\end{eqnarray*}}
\bibpunct{(}{)}{;}{a}{}{,} 

\usepackage{textcomp}
\newcommand{\foo}{\rlap{+}{\texttimes}}
\shorttitle{Modelling the cool loop system}
\shortauthors{Srivastava et al.}

\begin{document}
\title{Velocity Response of the Observed Explosive Events in the Lower Solar Atmosphere: I. Formation of the Flowing Cool Loop System}
\author{A.K.~Srivastava}
\affil{Department of Physics, Indian Institute of Technology (BHU), Varanasi-221005, India}
\author{Yamini~K.~Rao}, 
\affil{Department of Physics, Indian Institute of Technology (BHU), Varanasi-221005, India}
\author{P.~Konkol}
\affil{Group of Astrophysics, UMCS, ul. Radziszewskiego 10, 20-031, Lublin, Poland}
\author{K.~Murawski}
\affil{Group of Astrophysics, UMCS, ul. Radziszewskiego 10, 20-031, Lublin, Poland}
\author{M.~Mathioudakis}
\affil{Center of Astronomy and Physics, Department of Mathematics and Physics, Queen’s University, Belfast, UK}
\author{Sanjiv K. Tiwari}
\affil{Lockheed Martin Solar and Astrophysics Laboratory, 3251 Hanover Street, Building 252, Palo Alto, CA 94304, USA}
\affil{Bay Area environmental Research Institute, NASA Research Park, Moffett Field, CA 94035, USA}
\author{E.~Scullion}
\affil{Department of Mathematics \& Information Sciences, Northumbria University, Newcastle Upon Tyne, NE1 8ST, UK}
\author{J.G.~Doyle}
\affil{Armagh Observatory and Planetarium, College Hill, Armagh 9DG 73H, N. Ireland}
\author{B.N.~Dwivedi}
\affil{Department of Physics, Indian Institute of Technology (BHU), Varanasi-221005, India}
%


\begin{abstract}
We observe plasma flows in cool loops using the Slit-Jaw Imager (SJI) onboard the Interface Region Imaging Spectrometer (IRIS). Huang et al. (2015) observed unusually broadened Si IV 1403 \AA~line profiles at the footpoints of such loops that were attributed to signatures of explosive events (EEs). We have chosen one such uni-directional flowing cool loop system observed by IRIS where one of the footpoints is associated with significantly broadened Si IV line profiles. The line profile broadening indirectly indicates the occurrence of numerous EEs below the transition region (TR), while it directly infers a large velocity enhancement/perturbation further causing the plasma flows in the observed loop system. 
The observed features are implemented in a  model atmosphere in which a low-lying bi-polar magnetic field system is perturbed in the chromosphere by a velocity pulse with a maximum amplitude of 200 km s$^{-1}$. The data-driven 2-D numerical simulation shows that the plasma motions evolve in a similar manner as observed by IRIS in the form of flowing plasma filling the skeleton of a cool loop system. We compare the spatio-temporal evolution of the cool loop system in the framework of our model with the observations, and conclude that their formation is mostly associated with the velocity response of the transient energy release above their footpoints in the chromosphere/TR. Our observations and modeling results suggest that the velocity responses most likely associated to the EEs could be one of the main candidates for the dynamics and energetics of the flowing cool loop systems in the lower solar atmosphere. 
\end{abstract}

%

\section{Introduction}
The solar chromosphere comprises of complex magnetic structuring in active and quiet regions \citep[{\it e.g.},][and references cited there]{2014A&ARv..22...78W, 2019LRSP...16....1B} and  offers the opportunity of studying a variety of plasma motions and magnetic waves \citep[{\it e.g.},][and references cited there]{2004Natur.430..536D, 2007Sci...318.1591S, 2011Sci...331...55D, 2012Natur.486..505W, 2014Sci...346A.315T, 2014Sci...346D.315D, 2017NatSR...743147S, 2017Sci...356.1269M, 2018NatAs...2..951S}. These dynamical phenomena result in the transport of energy and mass from the lower solar atmosphere to the overlying corona contributing to heating and the origin of the nascent solar wind \citep[{\it e.g.},][and references cited there]{2002A&A...396..255D, 2009Sci...323.1582J, 2014Sci...346A.315T, 2017NatSR...743147S, 2017Sci...356.1269M, 2018NatAs...2..951S}. 

The curved magnetic field lines anchored in the concentration of small scale polarities at the quiet solar photosphere or the boundary/core of active regions, are most likely associated with the flowing material either bi-directional or from one footpoint to another of a low-lying loop system \citep[{\it e.g.},][and references cited there]{2004A&A...427.1065T, 2015ApJ...810...46H, 2019ApJ...887...56T,2019ApJ...874...56R}. Such plasma motions not only contribute to the mass cycle of the closed field corona (e.g., steady TR/coronal loops), but also serve to transport mass in the  nascent solar wind in the case of open magnetic channels \citep[{\it e.g.},][and references cited there]{2008A&A...478..915T,2008ApJ...676L.147H, 2009ApJ...704..883T}. Over the past three decades several studies have focused on plasma flows in hot active region coronal loops (AR loops), coronal holes (CHs),  quiet-Sun, and coronal hole boundaries in order to explore the inherent physical processes (e.g., waves, small-scale reconnection and nano-flares, etc) that create the mass/energy transport in such systems \citep[{\it e.g.},][and references cited there]{1993ApJ...411..406D, 1999ApJ...522L..77P, 2011A&A...534A..90D, 2014LRSP...11....4R, 2015SoPh..290.2889K}. 
\begin{figure*}
\centering
\hspace{-2cm}
\includegraphics[trim=2.5cm 5.0cm 5.0cm 0.5cm,scale=0.7,angle=90]{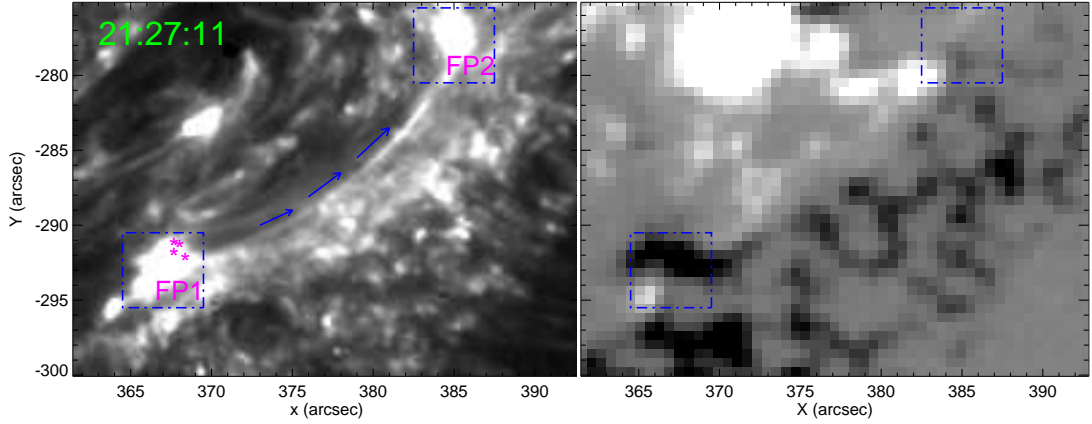}
\caption{Left: A cool loop system as seen in the IRIS SJI 1400 \AA~. Right: The magnetic field distribution at loop footpoints from SDO/HMI.}
\label{1}
\end{figure*}
The solar atmosphere exhibits downflows in the transition region (TR) which may give an evidence of coronal mass condensations as well as predominant cooling \citep[{\it e.g.},][and references cited there]{1999A&A...349..636T, 2003A&A...411..605M, 2008A&A...491L..13C, 2011ApJ...743..165F}.

Outward plasma flows in the TR and corona are mostly found to be associated with transient small-scale energy releases \citep[{\it e.g.},][and references cited there]{2003A&A...398..775M, 2007ApJ...661L.207W, 2013SoPh..282..453I, 2014SoPh..289.2971P, 2019ApJ...873...79C}.  Using IRIS observations, \citet{2015ApJ...810...46H} have found the presence of unusually large broadening in Si IV 1403\AA~directly at the footpoint of the cool and low-lying loop system. This broadening has been attributed to signatures of explosive events occurring at the footpoint of the cool loop system, and as a result, such energy release may help transport of plasma in the loop threads. In the present work, we consider the observations of \citet{2015ApJ...810...46H} and implement them in our model in the form of velocity perturbations which may be associated with the EEs below the TR. The 2-D model atmosphere is permeated by closed field lines and is structured vertically with the appropriate temperature profile, stratification and initial equilibrium. The model reproduces the flowing cool plasma in a curved magnetic field geometry, with observed fine structures, and shows that the cool loop systems are generated by the impulsive energy release and associated velocity perturbations at top of the chromosphere. In Sect.~2, we describe the obaservational analyses and results. A model of the impulsive plasma flows forming the cool loop system is given in Sect.~3. The numerical results are depicted in Sect.~4. Last section outlines the discussion and conclusions of the paper.

 \begin{figure*}
\centering
\hspace{-3cm}
\includegraphics[trim=2.5cm 5.0cm 5.0cm 0.5cm,scale=0.7,angle=90]{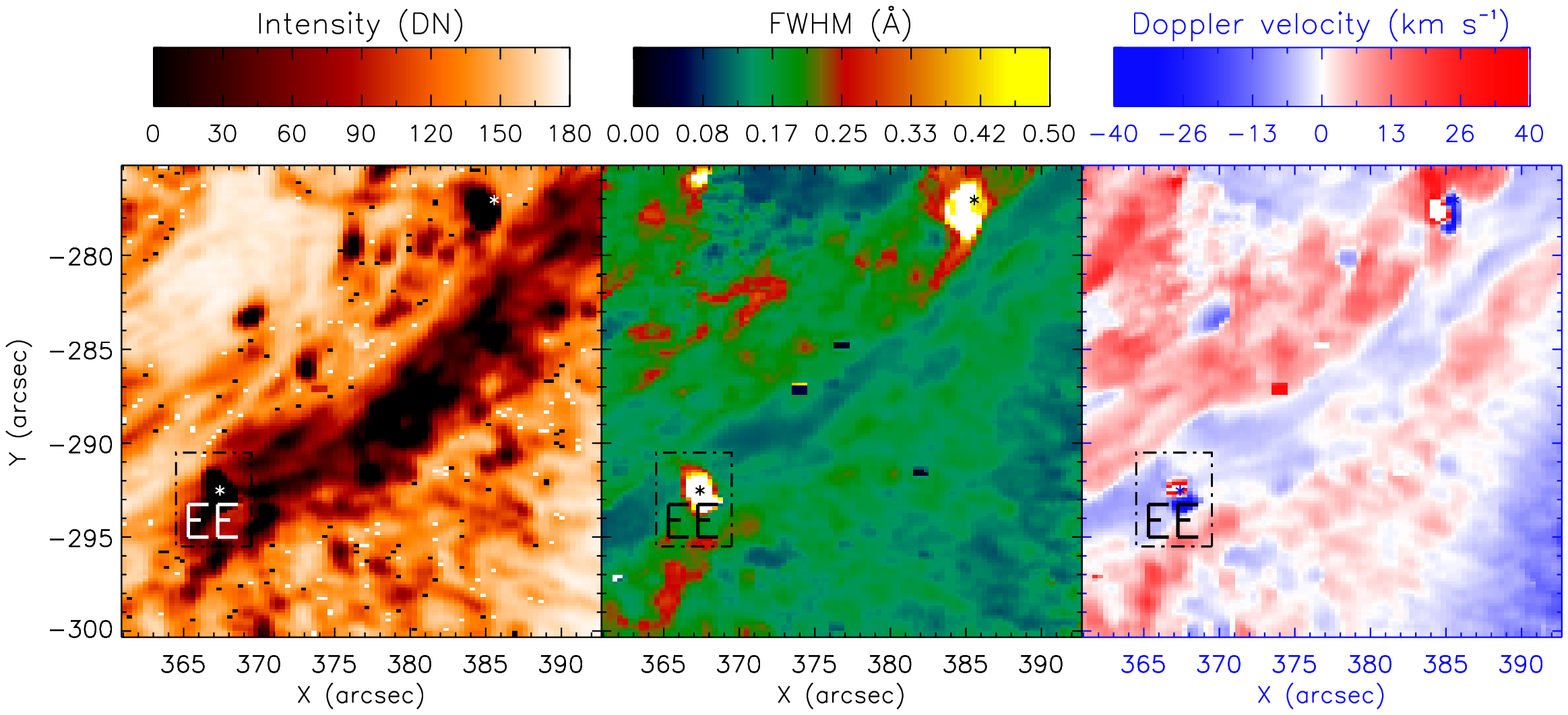}
\caption{Intensity (left), FWHM (middle), and Doppler velocity (right) maps derived from the Si IV~1403 \AA~line showing the evolution of the cool loop system.}
\label{2}
\end{figure*}
\section{Observational Analyses and Results}\label{obs_data}

The IRIS records spectra in the wavelength ranges of 1332 \AA~to 1358 \AA, 1389 \AA~to 1407 \AA~(Near Ultraviolet domain) and 2783 \AA~to 2834 \AA~(Far Ultraviolet domain).
The slit-jaw imager (SJI) provides context data that can help with the interpretation of the spectral rasters \citep{2014SoPh..289.2733D}.
The raster data used for our analysis were obtained in the active region AR 11934 on December 27, 2013 from 21:02:38 to 21:36:29 UTC. The slit covers a cool loop system during the whole raster scan. Very large dense raster spectral data has been used with the field of view 141\arcsec~ in the x-direction and 174\arcsec~in the y-direction having 400 steps of size 0.35\arcsec~with the step cadence of 5.1 s. The cool loop system is visible in the near ultraviolet (NUV) as well as the far ultraviolet (FUV) slit-jaw images. The IRIS Level 2 data which have already been corrected for flat fielding, dark current, and geometrical corrections have been used in our analysis.

Magnetic field information were obtained using data from the Helioseismic and Magnetic Imager (HMI) onboard Solar Dynamics Observatory (SDO). The HMI data has a pixel size of 0.5\arcsec \citep{2012SoPh..275..207S}.
The left panel of Fig.~\ref{1} shows the cool loop system in the slit-jaw image of 1400 \AA~line at 21:27:11 UTC corresponding to Si IV 1403 \AA~line in the spectral dataset. Different threads of the loops arepresent in the whole system. The FP1 and FP2 are two locations at the footpoints of the loop system that were further investigated. The right panel of Fig.~\ref{1} indicates the mixed magnetic polarities at the opposite footpoints of the loop system. This provides a chance for the occurrence of explosive events (EEs) and energy release above these footpoints \citep[e.g.]{2019ApJ...887...56T}. Both footpoints of the cool loop system are the active footpoints where bunch of flowing cool loop threads are anchored \citep{2015ApJ...810...46H}. Some loop threads flow from the first footpoint (FP1) to the second footpoint (FP2) where the activity is dominated above FP1 in terms of velocity enhancements/perturbations. While, in some other loop threads, the plasma flows in the opposite direction from FP2 to FP1. This is the reason that we observe the mixed flow patterns at both active footpoints in the chosen area FP1 and FP2. The physical process remains the same when we consider any typical flowing cool loop where the plasma is driven by the velocity enhancement from its one footpoint (FP1) to the other footpoint (FP2). We track the evolution of such plasma flows along a bunch of cool loop threads for our analysis where flow occurs from FP1 to FP2 (cf., Fig.~1, left-panel). The flow along the strand has been indicated by the blue arrows from FP1 to FP2.
We have simulated a specific part of the cool loop system where the plasma flows in one direction from FP1 to FP2. 
\begin{figure*}
\centering
\mbox{
\includegraphics[trim=0.5cm 0.5cm 0.2cm 0.2cm,scale=0.4,angle=90]{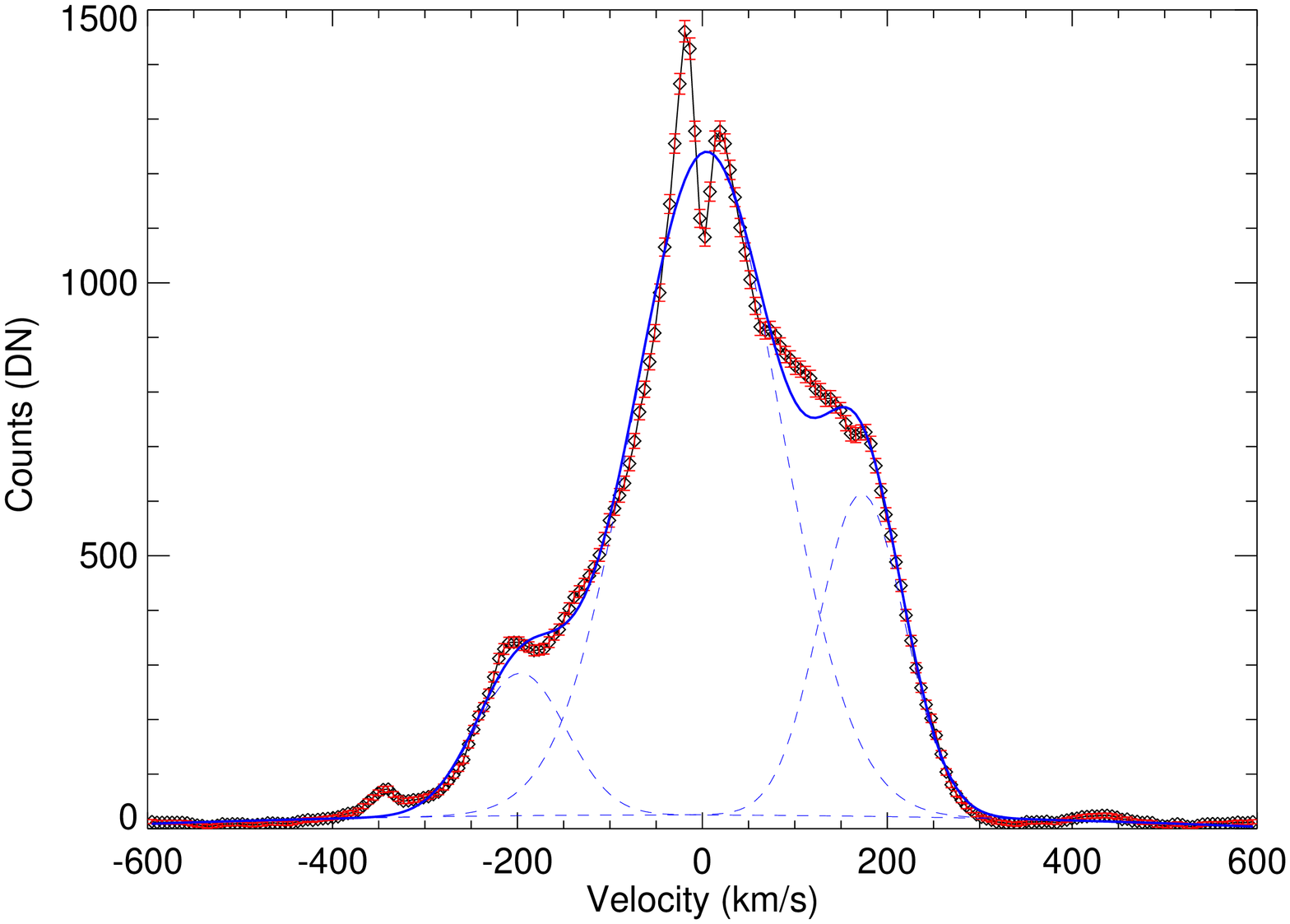}
\includegraphics[trim=0.5cm 0.5cm 0.2cm 0.2cm,scale=0.4,angle=90]{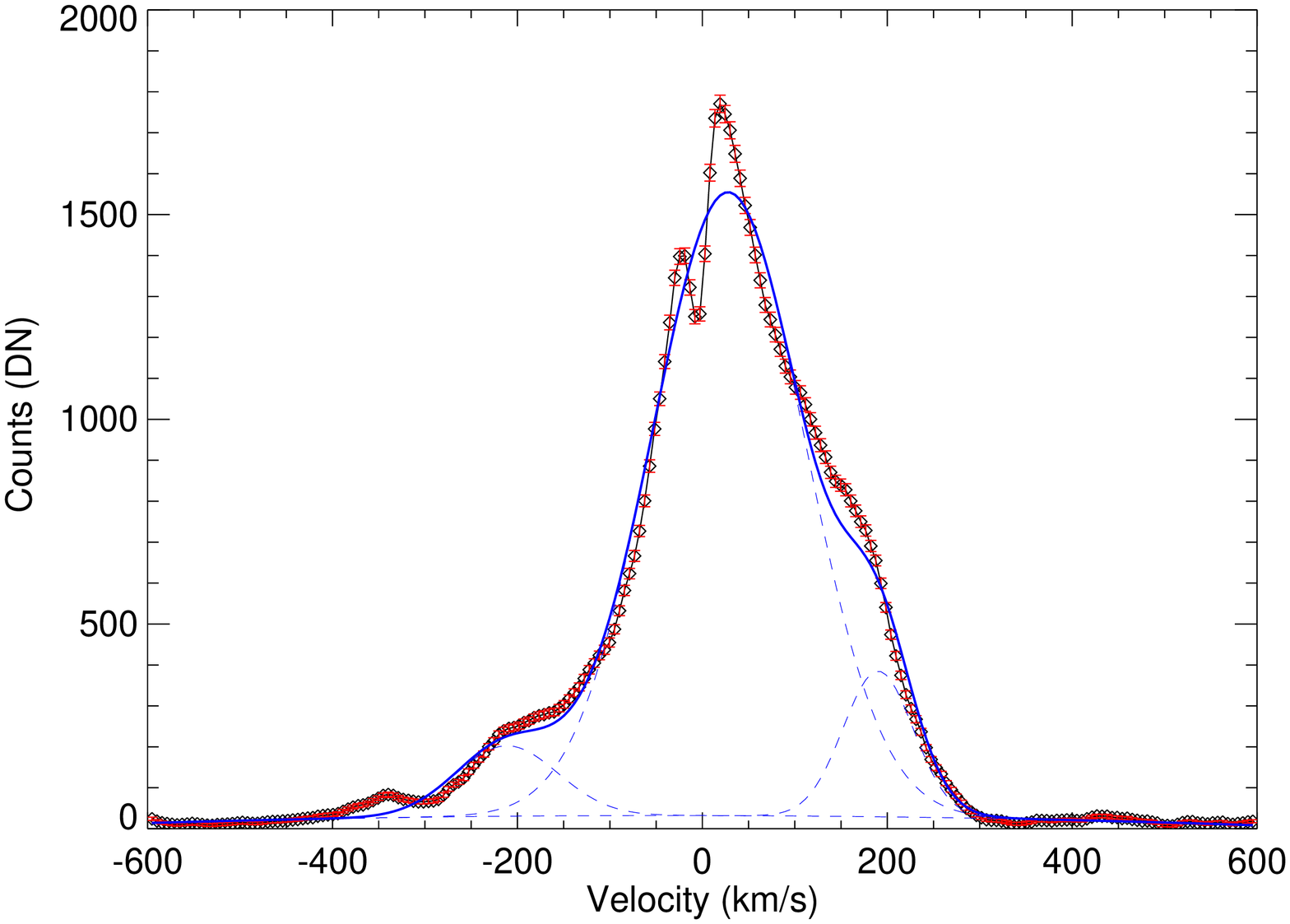}}
\mbox{
\includegraphics[trim=0.5cm 0.5cm 0.2cm 0.2cm,scale=0.4,angle=90]{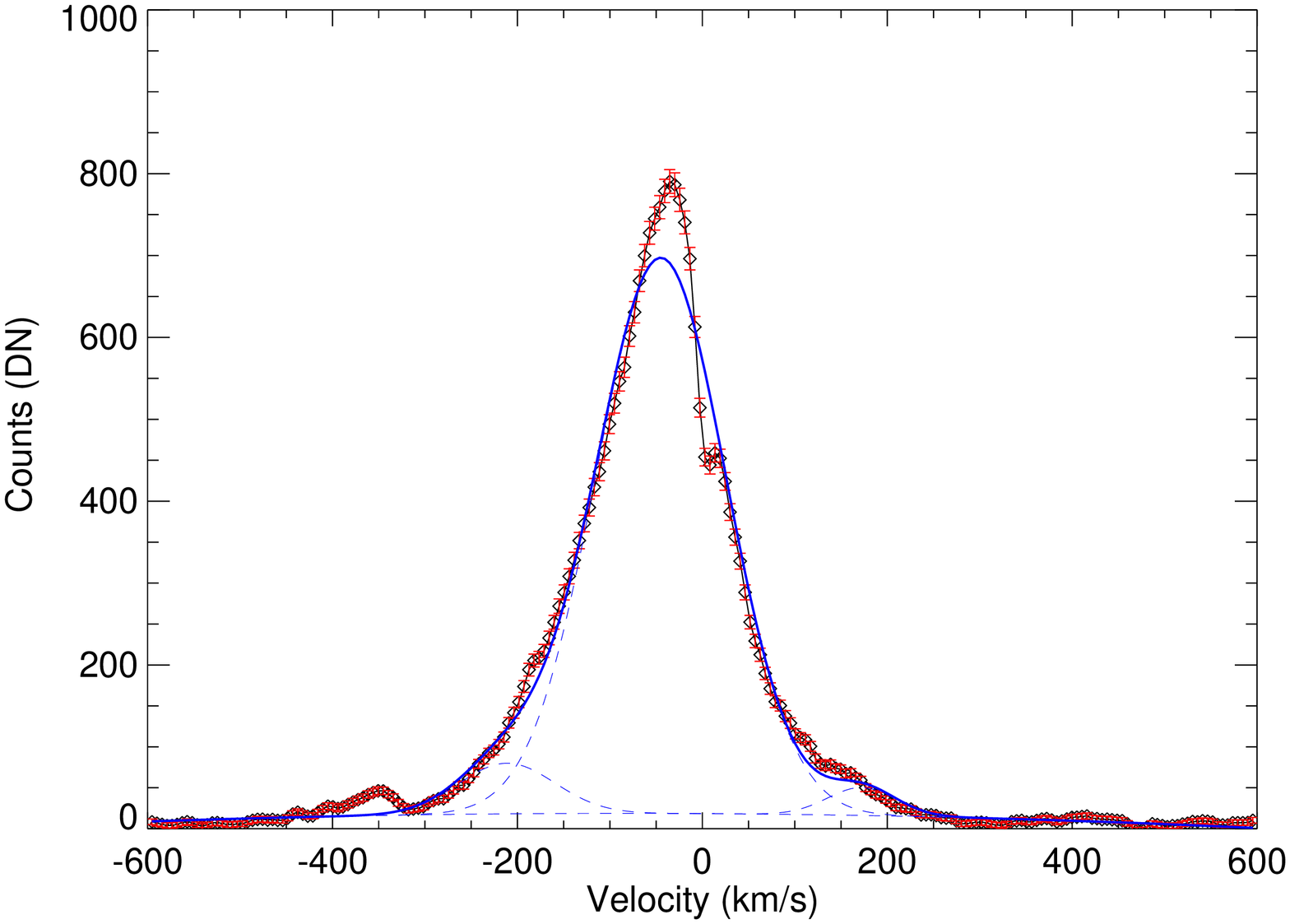}
\includegraphics[trim=0.5cm 0.5cm 0.2cm 0.2cm,scale=0.4,angle=90]{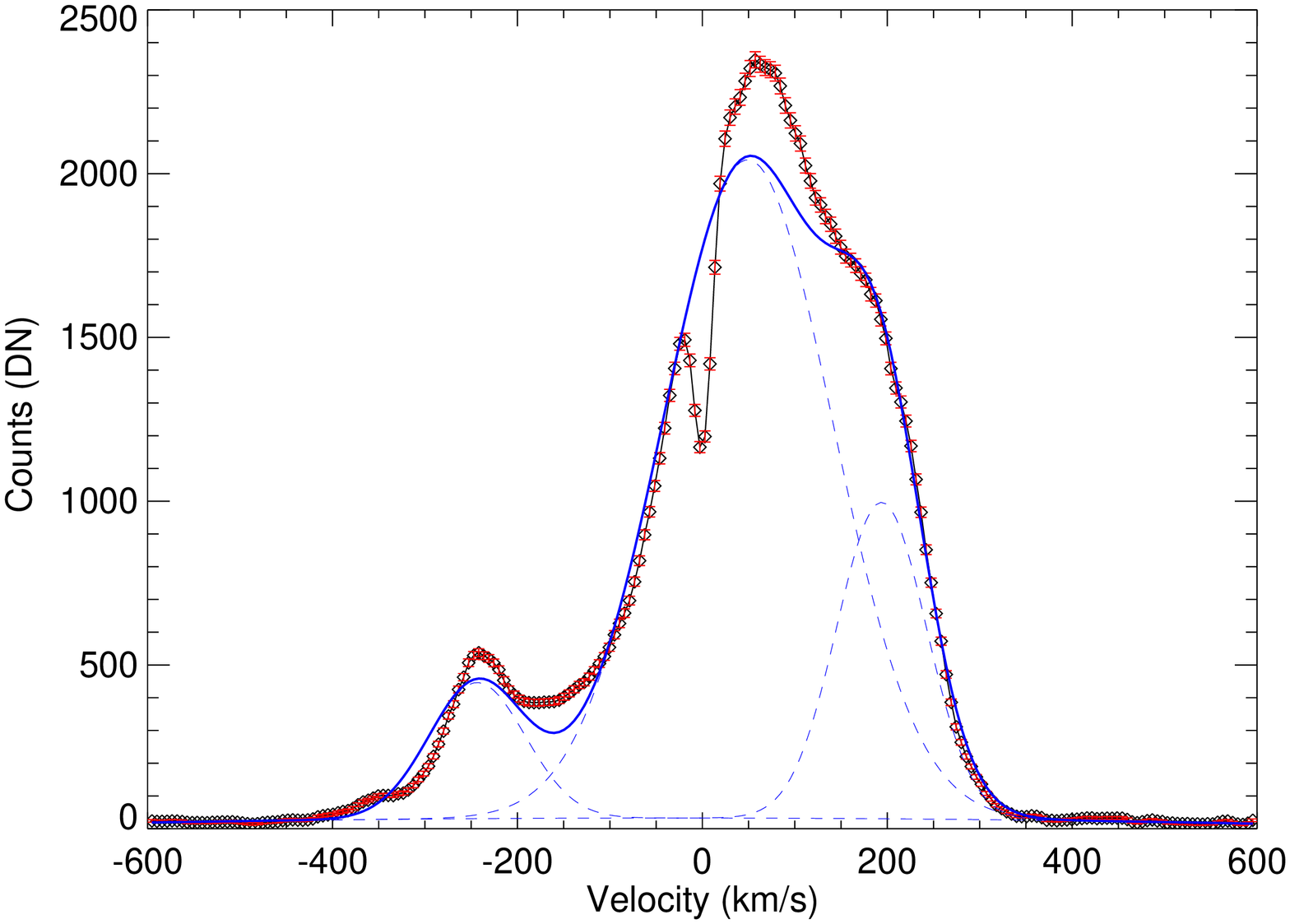}
}
\caption{The representative enhanced line-profiles of Si IV~1403 \AA~line at 4 different pixels near the EE location showing a velocity enhancement of around $\approx$200 km s$^{-1}$. Three individual dashed lines show the single Gaussian profiles used to fit the wing-enhancements.}
\label{3}
\end{figure*}
Fig.~\ref{2} shows the intensity, Doppler velocity and FWHM maps for the Si IV line. We have generated these maps by fitting a single Gaussian on the spectral profiles derived from each pixel of the chosen field-of-view. 
Plasma flows are present in both directions in different loop strands of the system exhibiting the mixed flow pattern. The box at the blue-shifted footpoint (FP1) is overlaid where the marked pixel locations are used to derive the unusually broadened spectral profiles.

\begin{figure*}
\centering
\includegraphics[scale=0.6,angle=90,width=8cm,height=8cm,keepaspectratio]{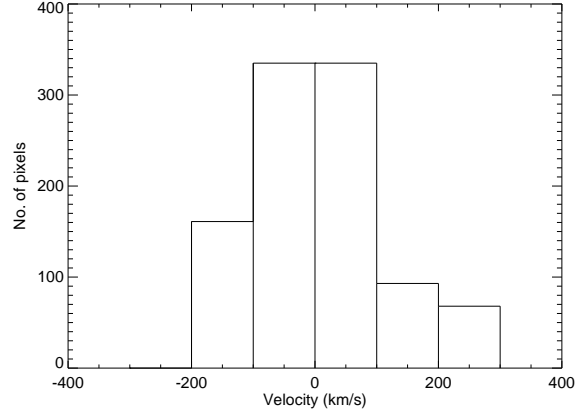}

\caption{The histogram showing the distribution of maximum enhanced velocity over the wing of the observed line profiles.}
\label{4}
\end{figure*}

Fig.~\ref{3} shows a few example profiles near the location marked by an astrisk symbol where enhanced line widths out to $\pm$200 km s$^{-1}$ can be seen. Multiple locations are shown in Fig.~\ref{1} by astrisk. These locations may be associated with EEs \citep{2015ApJ...810...46H}. However, we observe only the velocity response of such possible EEs in the TR in the present observational base-line. Approximately 7 $\%$ of the individual pixels within the box in Fig.~\ref{2} have broadened line-profiles of Si IV for which the distribution is shown in Fig.~\ref{4}. The dotted blue lines in Fig.~3 show three different Gaussian profiles required to fit the broadened line profile. The solid blue line shows the fitted profile.
We have chosen different pixel locations  above the left footpoint of the cool loop system compared to \citet{2015ApJ...810...46H} where we also find the enhanced line-broadening at multiple locations. 

\section{A model of the impulsive plasma flows forming the cool loop system} \label{model}

In order to model the chromospheric cool plasma flows in the curved magnetic loop system we take into account a gravitationally-stratified and magnetized 2-D solar atmosphere that is described by the set of ideal MHD equations as (\citealt{2007ApJS..170..228M}; \citealt{2012ApJS..198....7M}, \citealt{WMMM2014}; \citealt{angeo-2019-67})
\begin{eqnarray} 
    \frac{\partial \rho}{\partial t} + \nabla \cdot \left( \rho \mathbf{v} \right) &=& 0 \,,
        \label{eq:mhd:1} \\
    \rho \frac{\partial \mathbf{v}}{\partial t} + \rho \left( \mathbf{v} \cdot \nabla \right) \mathbf{v} &=&
        - \nabla p + \frac{1}{\mu} \left( \nabla \times \mathbf{B} \right) \times \mathbf{B} + \rho \mathbf{g} \,,
        \label{eq:mhd:2} \\
    \frac{\partial \mathbf{B}}{\partial t} &=& \nabla  \times \left( \mathbf{v} \times \mathbf{B} \right) \,,
        \label{eq:mhd:3} \\
    \nabla \cdot \mathbf{B} &=& 0 \,,
        \label{eq:mhd:4} \\
    \frac{\partial p}{\partial t} + \nabla \cdot \left( p \mathbf{v} \right) &=&
        \left( 1 - \gamma \right) p \nabla \cdot \mathbf{v} \,,
        \label{eq:mhd:5} \\
    p &=& \frac{k_{\mathrm{B}}}{m} \rho T \,
        \label{eq:mhd:6}
\end{eqnarray}
, where $\mathbf{v}$ depicts the velocity field, $\mathbf{B}$ is the magnetic field satisfying the divergence-free condition at each and every point of the solar atmosphere, $\mu$ defines the magnetic permeability, $p$ is the thermal pressure, and $B^2/(2\mu)$ is magnetic pressure. 
The specific heat ratio $\gamma$ is chosen as $5/3$. 
The symbol $\rho$ denotes the mass density. $T$ represents the temperature and $k_{\mathrm{B}}$ is the Boltzmann constant. The symbol $m$ depicts the mean particle mass in the ionized plasma. 
We consider a typical value of $m$ to be $1.24$ for the model atmosphere. For the sake of simplicity, we consider the ideal conditions and ignore non-ideal effects such as viscosity and magnetic diffusivity, radiative cooling and/or heating of the plasma, as we are primarily interested in understanding the kinematics and evolutionary properties of the cool loop system.

\begin{figure*}
\mbox{
\includegraphics[scale=0.4,angle=0,width=8cm,height=10cm,keepaspectratio]{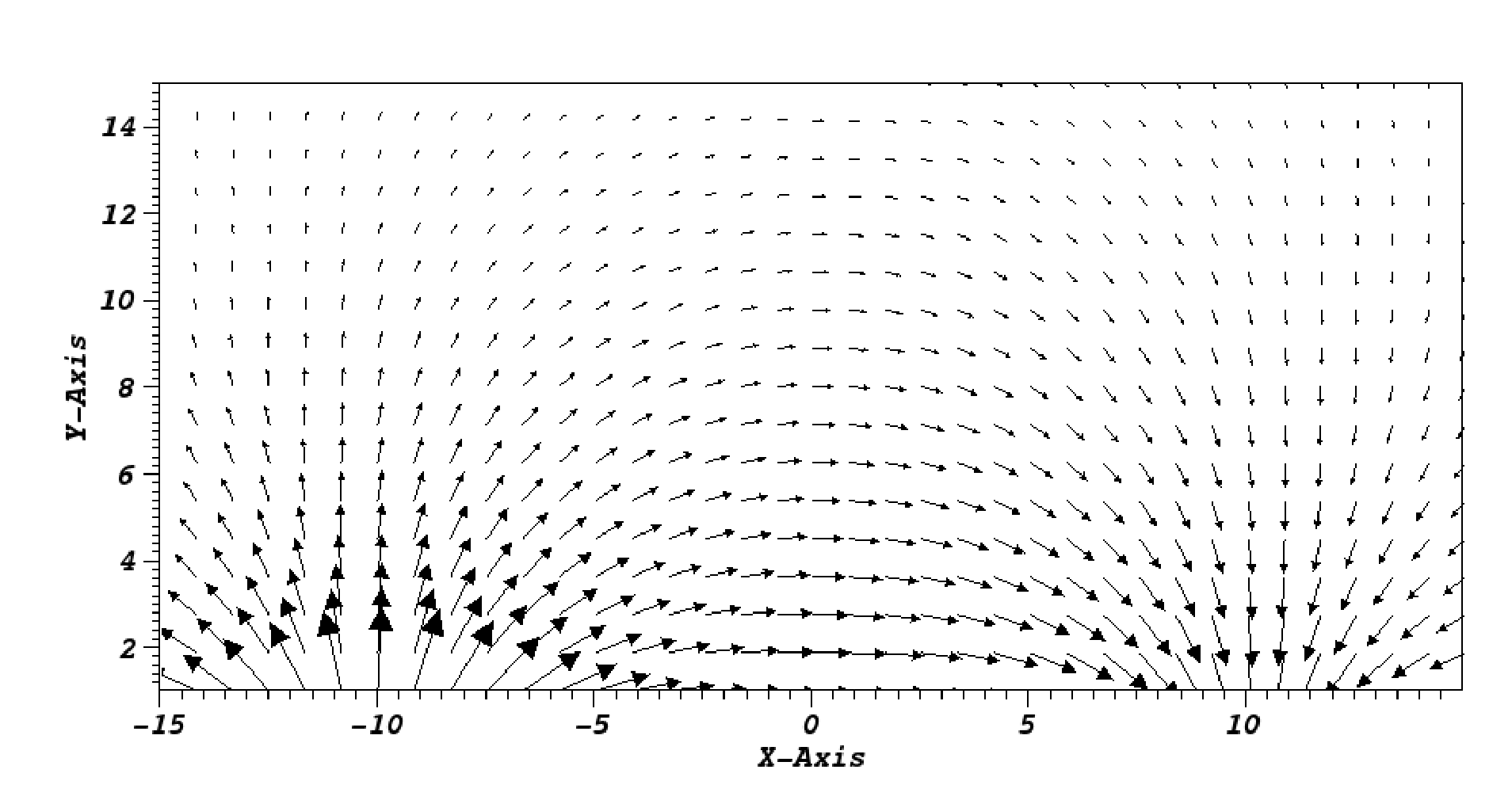}
\includegraphics[scale=0.4,angle=0,width=8cm,height=8cm,keepaspectratio]{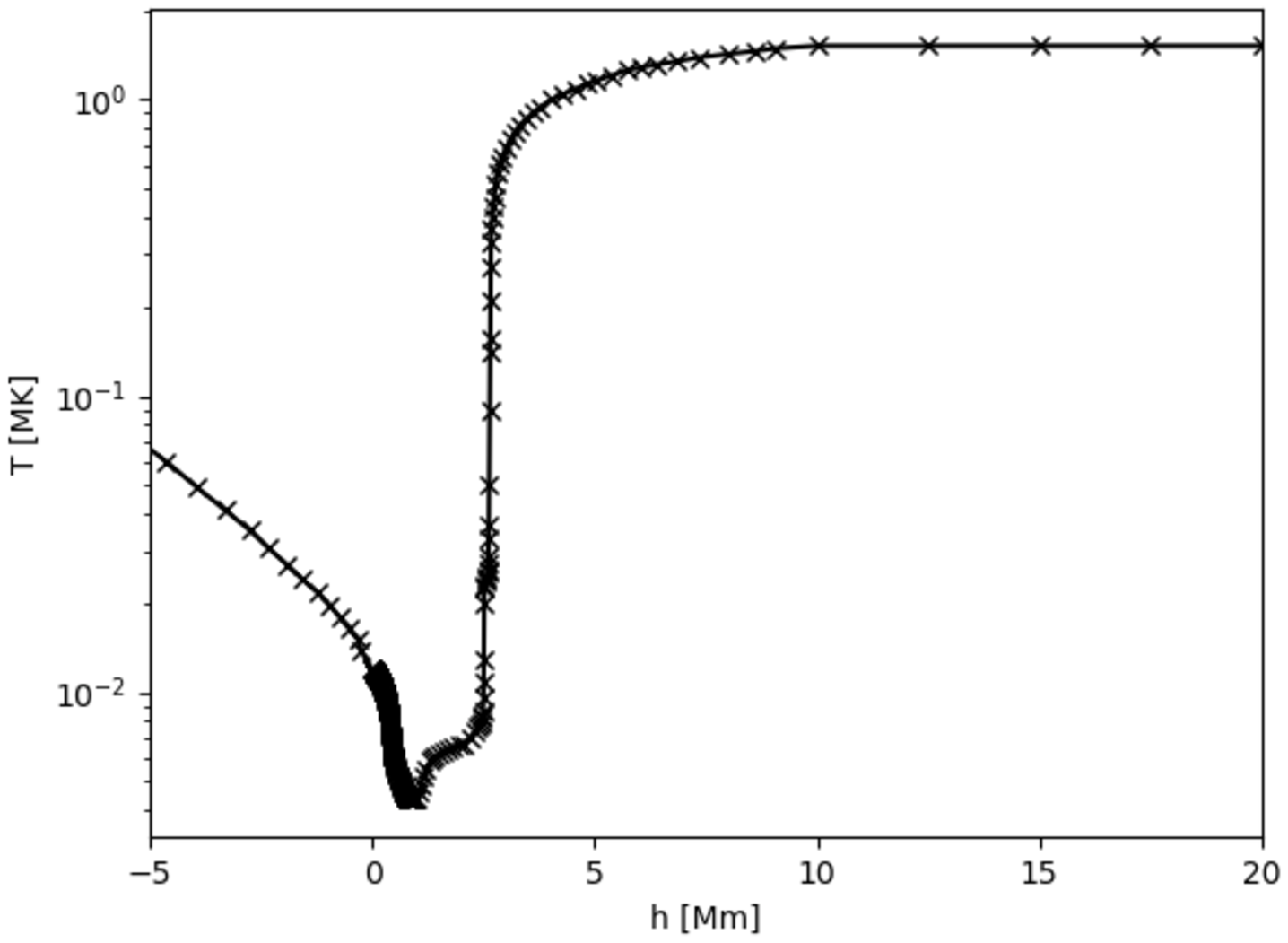}
}
\caption{Left-panel: Equilibrium bipolar magnetic field vectors in the model solar atmosphere. X-axis and Y-axis are given in Mm. Right-panel: Temperature profile derived from the model of \cite{2008ApJS..175..229A}.}
\label{mag_field}
\end{figure*}


\subsection{Initial conditions}
We take the stationary state of MHD equations as the basis for our initial conditions. By putting time derivatives to zero ${\partial}/{\partial t} = 0$ and assuming $\mathbf{v}=\mathbf{0}$, the only nontrivial conclusion from equation (\ref{eq:mhd:2}) takes the form:
\begin{equation}
     \mathbf{0}= \frac{1}{\mu} \left( \nabla \times \mathbf{B}_\mathrm{e} \right) \times \mathbf{B}_\mathrm{e}
         - \nabla p_\mathrm{e} + \rho_\mathrm{e} \mathbf{g} \,. \label{eq:mhd:equal}
\end{equation}

where the subscripts "$\mathrm{e}$" represents the equilibrium quantities.

By choosing current-free magnetic field, we can further split equation (\ref{eq:mhd:equal}) into
\begin{eqnarray} 
    \mathbf{\nabla} \times \mathbf{B_\mathrm{e}} &=& \mathbf{0} \,,
        \label{eq:mhd:equal:1} \\
    -\nabla p_\mathrm{e} + \rho_\mathrm{e} \mathbf{g} &=& \mathbf{0} \,.
        \label{eq:mhd:equal:2}
\end{eqnarray}
The current-free magnetic field (\ref{eq:mhd:equal:1}) has always corresponding scalar field $\phi_\mathrm{e}(x, y)$, and we can deduce the magnetic flux function given as follows \citep{2010A&A...521A..34K}:
\begin{equation}
  \mathbf{B_\mathrm{e}}(x, y) = \nabla \phi_\mathrm{e} = [B_{\mathrm{e}x}, B_{\mathrm{e}y}, 0] =
    \nabla \times (A_\mathrm{e}\widehat{z}) 
\end{equation}
with,
\begin{equation}
    \phi_\mathrm{e}(x, y) = \frac{S_1(x_2 - a_{11})}{(x_1 - a_{11})^2 + (x_2 - a_{12})^2} + 
                            \frac{S_2(x_2 - a_{21})}{(x_1 - a_{21})^2 + (x_2 - a_{22})^2} \,,
    \label{eq:mhd:equal:scalar:potential}
\end{equation}
\begin{equation}
    A_\mathrm{e}(x, y) = \frac{S_1(x_1 - a_{11})}{(x_1 - a_{11})^2 + (x_2 - a_{12})^2} + 
                         \frac{S_2(x_1 - a_{21})}{(x_1 - a_{21})^2 + (x_2 - a_{22})^2} \,.
    \label{eq:mhd:equal:vector:potential}
\end{equation}

Where $S_1$ and $S_2$ are the strength of the two poles. The magnitude of two poles are taken as $\approx$330 Gauss, which is typical of the quiscent loop threads anchored in tiny magnetic polarities. The pair of variables $[a_{11}, a_{12}]$ and $[a_{21}, a_{22}]$ are their positions. $a_{11}$ and $a_{21}$ are the X-positions for the first and second poles respectively while $a_{12}$ and $a_{22}$ represent the vertical positions of the first and second poles respectively.The first subscript signifies the sequel of the pole ('1': first pole; '2': second pole). The second subscript depicts the axes ('1': X-axis; '2': Y-axis).
The resultant vector potential for the curved loop is obtained by adding the similar expression for the potential for two opposite poles of the loop system. 
The above mentioned expressions mimic the bipolar closed field lines representing the loop system. In the above mentioned expression, we fix the vertical coordinate of magnetic poles, $a_{21} = a_{22} = -5$ Mm in the convection zone. Also, $a_{11} = -10$ Mm and $a_{11} = -a_{12}$. The magnetic field lines exhibit the curved and low lying loop system as shown in Fig.~\ref{mag_field}. We take the magnetic field strength typical of the regions as measured at the footpoint of the cool loop system (Fig.~\ref{1}; HMI observations).
Expressions (\ref{eq:mhd:equal:scalar:potential}) and (\ref{eq:mhd:equal:vector:potential}) may be considered as a transformation to the new orthogonal coordinate system with new coordinates $\phi_\mathrm{e}$ and $A_\mathrm{e}$.

The model solar atmosphere obeys the hydrostatic equilibrium which is represented as (cf., Eq.~9 also):\\
\begin{equation}
- \bigtriangledown p_\mathrm{e} + \rho_\mathrm{e} g = 0
\end{equation}
The value of g is kept fixed at 274 m $s^{-2}$. Using the vertical component of the hydrostatic equilibrium in the model solar atmosphere and the ideal gas law, we determine the equilibrium plasma gas pressure and mass density w.r.t. the vertical direction ($y$) as follows \citep{angeo-2019-67}:\\
\begin{equation}
p(y) = p_{ref} exp\left(- \int_{yref}^{y}\frac{dy'}{\Lambda(y')}\right),         \rho(y) = \frac{p(y)}{g\Lambda(y)}, 
\end{equation}
where,
\begin{equation}
\Lambda(y) =\frac{k_{B} T(y)}{\hat{m}g},
\end{equation}
where $\Lambda(y)$ is the pressure scale height, $k_{B}$ is the Boltzmann constant, T(y) is the temperature profile w.r.t height, $\hat{m}$ is the proton mass, and g is the gravitational acceleration.In Eq.~14, $p_{ref}$ is a gas pressure at the reference level $y_{ref}$.The reference level is taken in the overlying corona at $y_{ref}$ = 10 Mm.
\\
For considering the gravitationally stratified and vertically structured solar atmosphere, we obtain the plasma temperature profile $T(y)$ which is derived by \citet{2008ApJS..175..229A} as displayed in Fig.~\ref{mag_field} (right-panel). It should be noted that the typical value of the temperature $T(y)$ at the top of the photosphere is about 5700 K. This temperature corresponds to $y$ = 0.5 Mm and falls gradually attaining its minimum of 4350 K at $y$ = 0.95 Mm which represents the temperature minimum. As we move higher up in the atmosphere, $T(y)$ increases gradually up to the TR which is located at approximately y$ = 2.7 Mm$. $T(y)$ sharply increases up to the corona and finally attains the value of mega-Kelvin at the coronal heights as shown in Fig.~\ref{mag_field} (right panel).
This temperature model is also described in detail by \citet{angeo-2019-67}.

\subsubsection{Perturbations}
We perturb the hydrostatic equilibrium atmosphere by the initial pulse in the vertical component of the velocity. The Gaussian form of the velocity pulse in the vertical direction is given as follows:\\
\begin{equation}
V(x,y,t=0) = A_\mathrm{v} \times exp\left(-\frac{(x - x_{0})^2 + (y - y_{0})^2}{\omega^2}\right)\,.
\end{equation}
Here ($x_\mathrm{0}$, $y_\mathrm{0}$) is the initial position of the velocity pulse, $\omega$ is the width of the pulse, and $A_\mathrm{v}$ denotes the amplitude. We fix the value of $x_{0}$ = $-$8.6 Mm, $y_{0}$ = 2.0 Mm. Therefore, for launching the velocity perturbations in the chromosphere just below the TR, we take $\omega$  equals to 0.06 Mm. 
It should be noted that the magnitude of the velocity pulse A$_\mathrm{v}$ is taken as 200 km s$^{-1}$ as depicted by the enhanced line-profile of Si IV 1402.77 \AA~as shown in Fig.~\ref{3}, which has also been reported by \citet{2015ApJ...810...46H}.

\subsection{Numerical Methods}
We consider the numerical simulation box representing a realistic solar atmosphere covering the region from the photosphere to the inner corona where curved magnetic field lines of the loop system are present (cf., Fig.~\ref{mag_field}, left-panel). In the solar atmosphere maintained at the hydrostatic equilibrium, the realistic temperature profile T($y$) is shown in Fig.~\ref{mag_field} (right-panel), which is taken from the measurement of \cite{2008ApJS..175..229A}, and also reported in \citet{2010A&A...521A..34K} and \citet{angeo-2019-67}.
The PLUTO code  is a Godunov-type, non-linear, finite-volume magnetohydrodynamic (MHD) code which takes into account both ideal and non-ideal sets of governing equations (cf., \citealt{2007ApJS..170..228M}, \citealt{2012ApJS..198....7M}, \citealt{WMMM2014}). It is constructed to integrate numerically a system of conservation laws which can be shown as follows: \\ 
\begin{equation}
\frac{\partial U}{\partial t}\ + \nabla \cdot F(U) = S(U)\,.
\end{equation}
In this equation, U denotes a set of conservative physical fields (e.g., magnetic field, density, velocity, pressure etc.), while F(U) is the flux tensor and S(U) is the source term \citep{angeo-2019-67}.\\
In order to solve  the ideal MHD equations numerically, we set the simulation box at (-15, 15) Mm $\times$ (1, 15) Mm. This represents a region of the solar atmosphere of  30 Mm and 14 Mm that spans in the horizontal and vertical directions respectively. This solar atmosphere is constructed within the simulation box, with all four boundary conditions to their equilibrium values. 
The numerical simulations are carried out with double precision using multiple passage interface (MPI) (see \citealt{2007ApJS..170..228M}). Eight processors are used in the parallel calculations taking approximately 10 hours of CPU time for each set of the calculations. We have adopted a static uniform grid (-15 to 15 Mm) which is divided into 1500 equal cells in the x-direction. We have also implemented a static uniform (1 to 5 Mm) and stretched (5 to 15 Mm) grid divided into 200 and 500 cells respectively in the y-direction. The resolution of the simulation domain is 20 km per numerical cell. We have stored the simulation data every 10 seconds. \\
In our numerical modelling, we set the Courant-Friedrichs-Lewy number equals to 0.25. We use Roe solver for the flux computation, which is linearized Riemann solver based on the characteristic decomposition of the Roe matrix (cf., \citealt{2007ApJS..170..228M, 2012ApJS..198....7M}, \citealt{WMMM2014}, \citealt{angeo-2019-67}). 

\begin{figure*}[h]
\centering
\includegraphics[scale=0.4,angle=0,width=8cm,height=8cm,keepaspectratio]{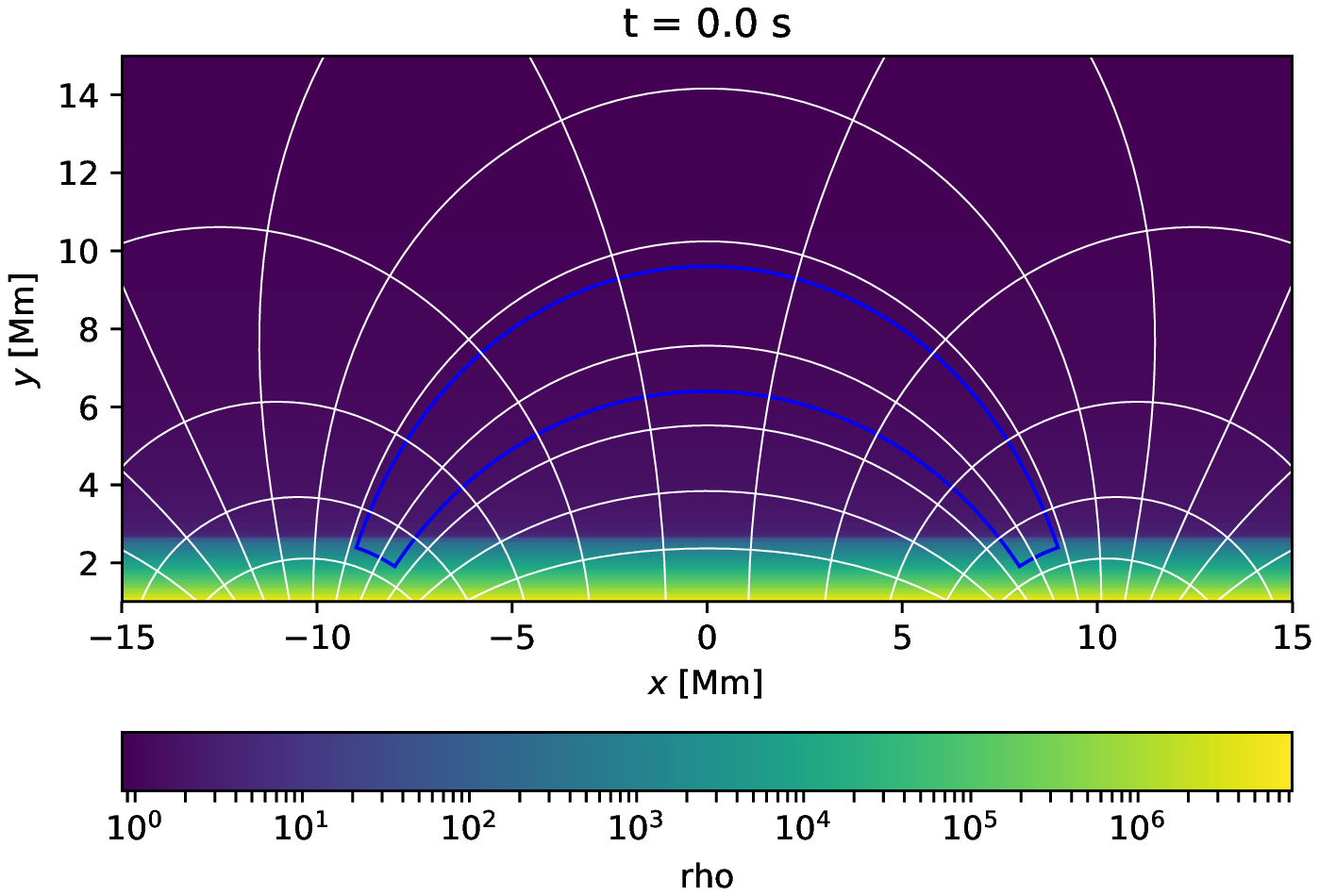}    
\\
\includegraphics[scale=0.4,angle=0,width=8cm,height=8cm,keepaspectratio]{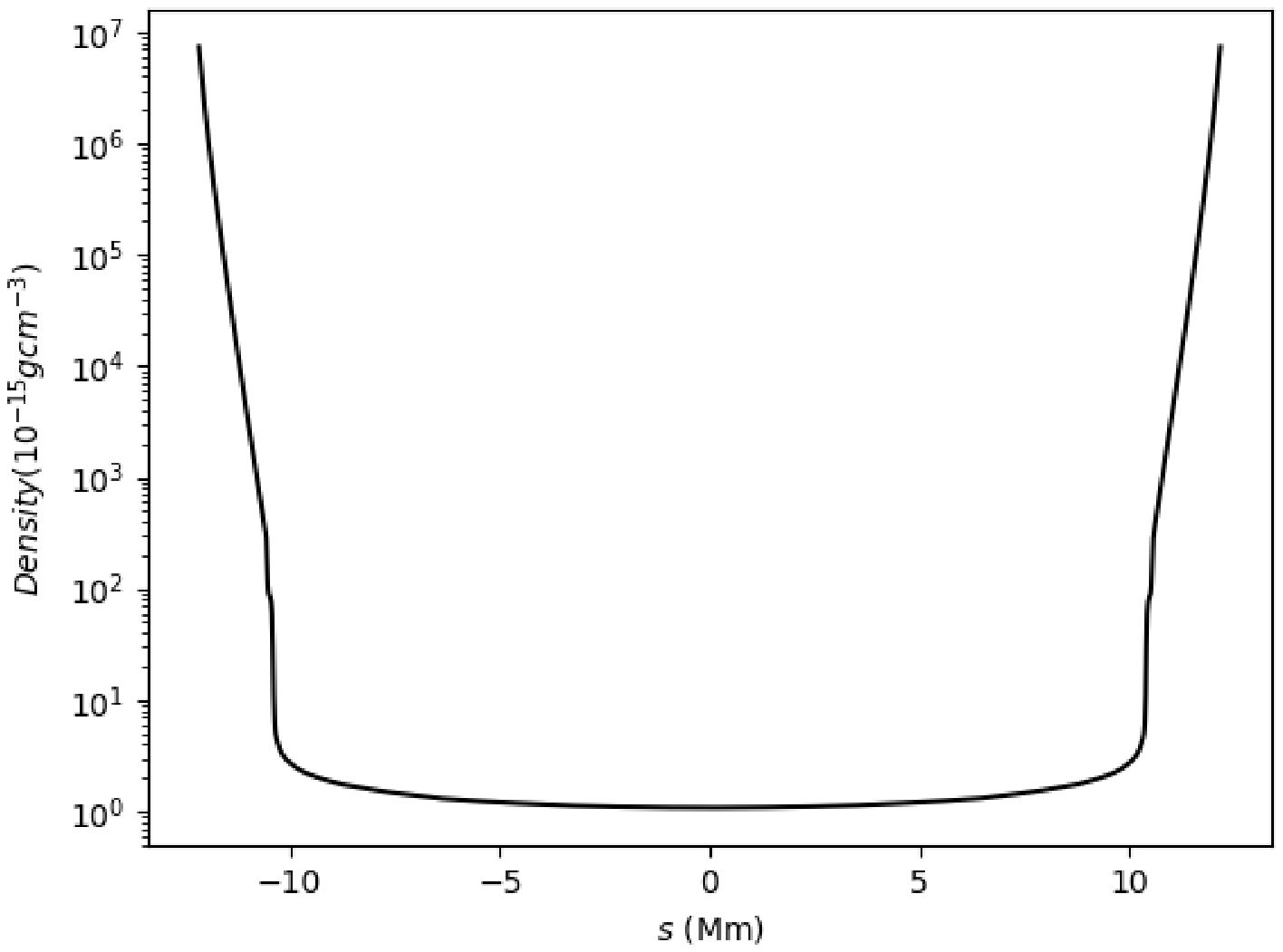}
\caption{Top-panel: Bipolar initial solar atmosphere. $\phi{_e}$=constant represnts the  equipotential line and A$_{ez}$=constant are lines of magnetic force, which are perpendicular to each other. All these lines are represented in white-colour. The low-lying bipolar loop system is shown within the blue-curved box area, where plasma flows fill the magnetic field lines. Initially mass is not filled and the transient cool-loop system is not formed there. Bottom-panel: Equilibrium normalized mass density profile along the chosen curved magnetic fields. The mass density is expressed in units of 10$^{-15}$ g cm$^{-3}$.}
\label{density_prof}
\end{figure*}

\begin{figure*}
\includegraphics[scale=0.5,angle=0,width=16cm,height=16cm,keepaspectratio]{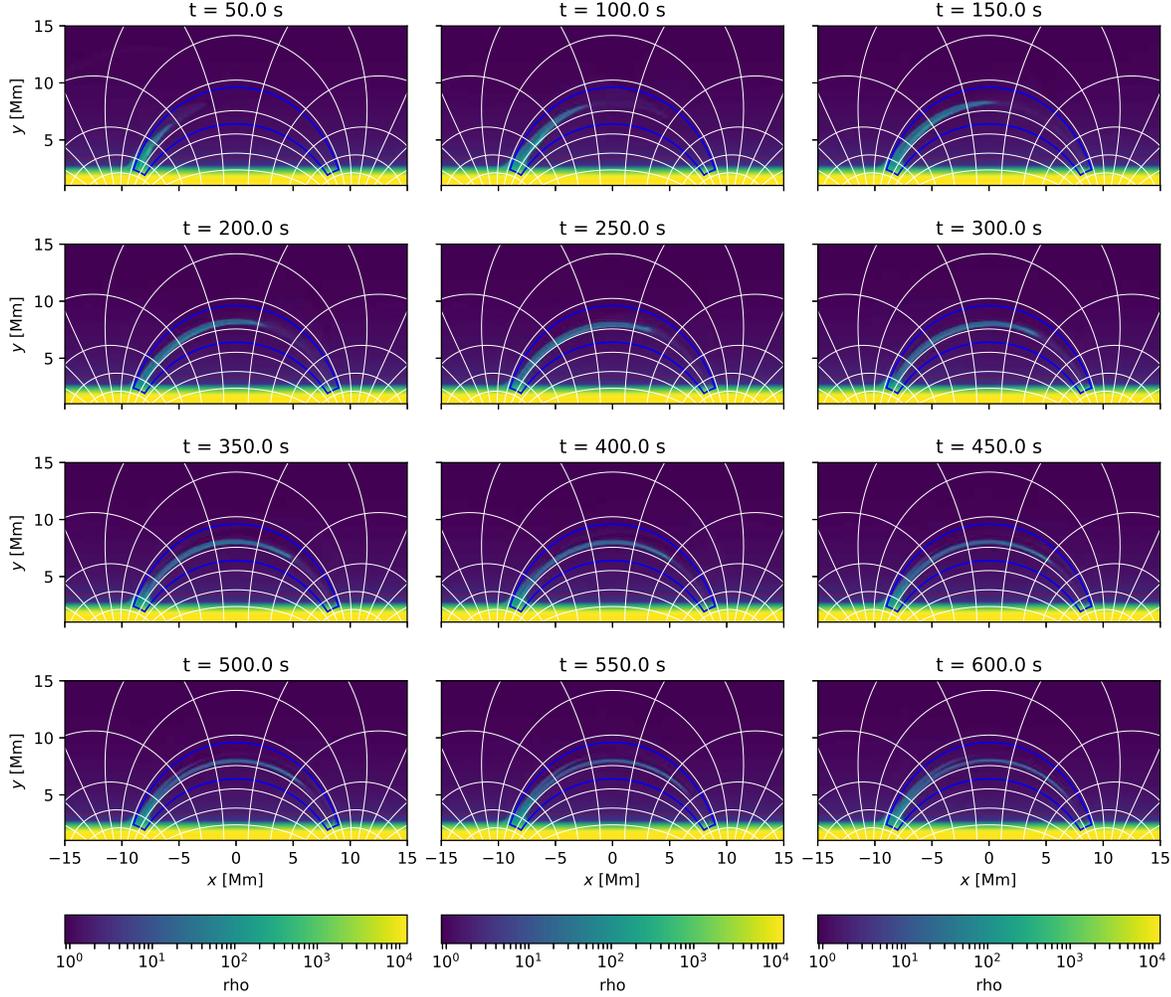}
\caption{Spatio-temporal evolution of the cool-loop system.}
\label{fig4}
\end{figure*}

\begin{figure*}
\includegraphics[scale=0.5,angle=0,width=16cm,height=16cm,keepaspectratio]{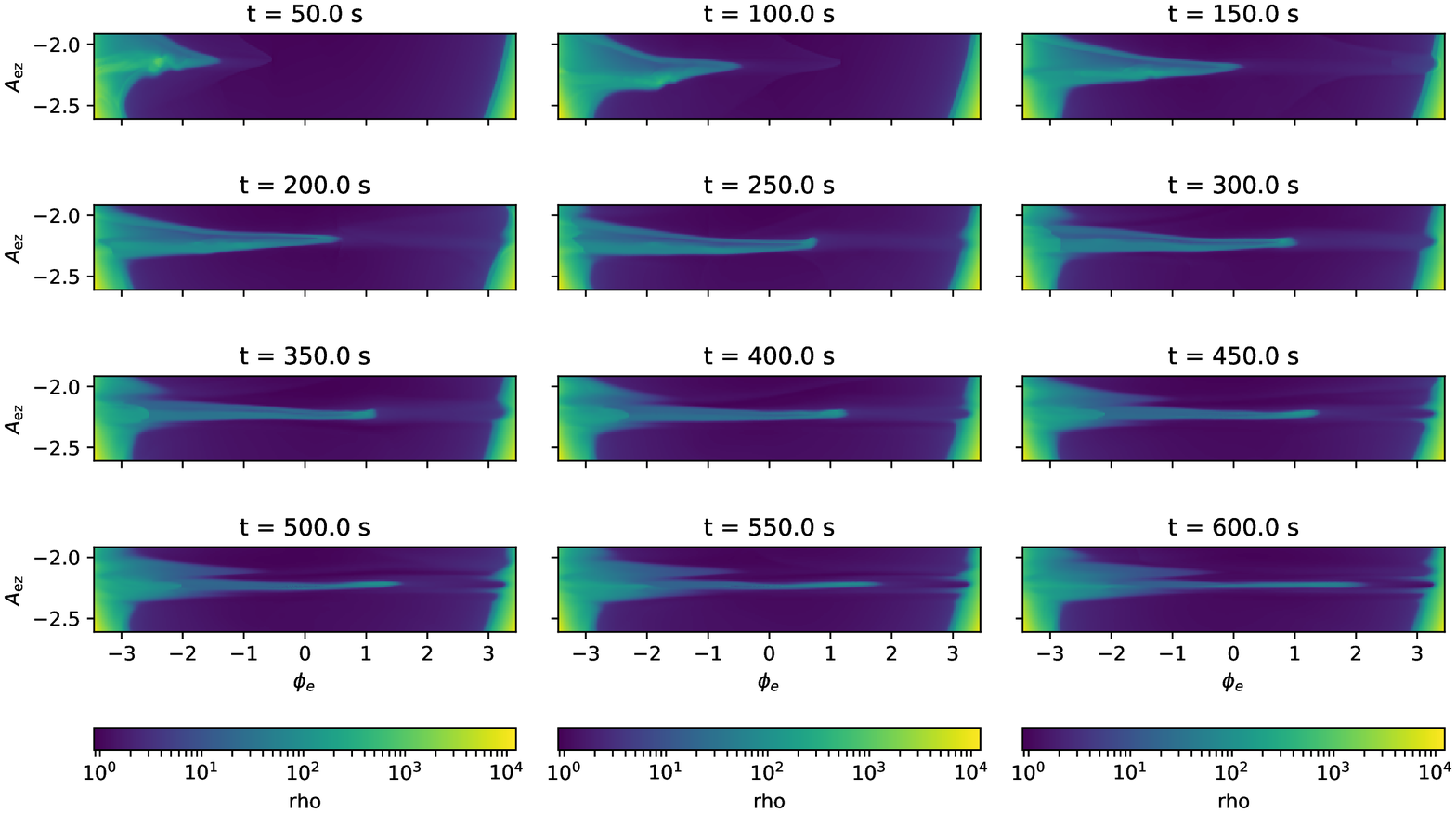}
\caption{A$_{z}$ vs $\phi$ representation of the cool-loop system providing the straight view of the plasma flows along the chosen set of the magnetic field skeleton of the cool loop. The evolution of the plasma flows and associated fine structures is clearly evident in it. Straightened region in this figure is equivalent to the region shown in the blue box in Figs.~6-7.}
\label{evolve_fine}
\end{figure*}

\section{Numerical Results}
Fig.~\ref{fig4} represents the spatio-temporal evolution of the cool plasma in the bipolar  magnetic field system as shown by the black curve in Fig.~\ref{density_prof} in various density maps. The density is expressed in units of $10^{-15}$ g cm$^{-3}$. 
The plasma flows along the magnetic field lines, and crosses the loop apex at t$\approx$300 s. This spatio-temporal evolution mimics the observed plasma flows in the cool loop system from the  left footpoint to the right (Fig.~\ref{1}, left panel; Movie 1). As seen in the observations (Movie 1), once the plasma moves towards the other footpoint, it tends to reflect back and oppositely encounter the up-flowing plasma and complex mass motions arise. In the observations (Movie 1), the temporal scale of the major mass flow in cool loop approximately occurrs for 10 minutes ($\sim$21:18-21:28 UT). Almost similar time-scale of the plasma flow is found in the model cool loop system of almost similar length.  
\begin{figure*}
\mbox{
\includegraphics[scale=0.4,angle=0,width=8cm,height=12cm,keepaspectratio]{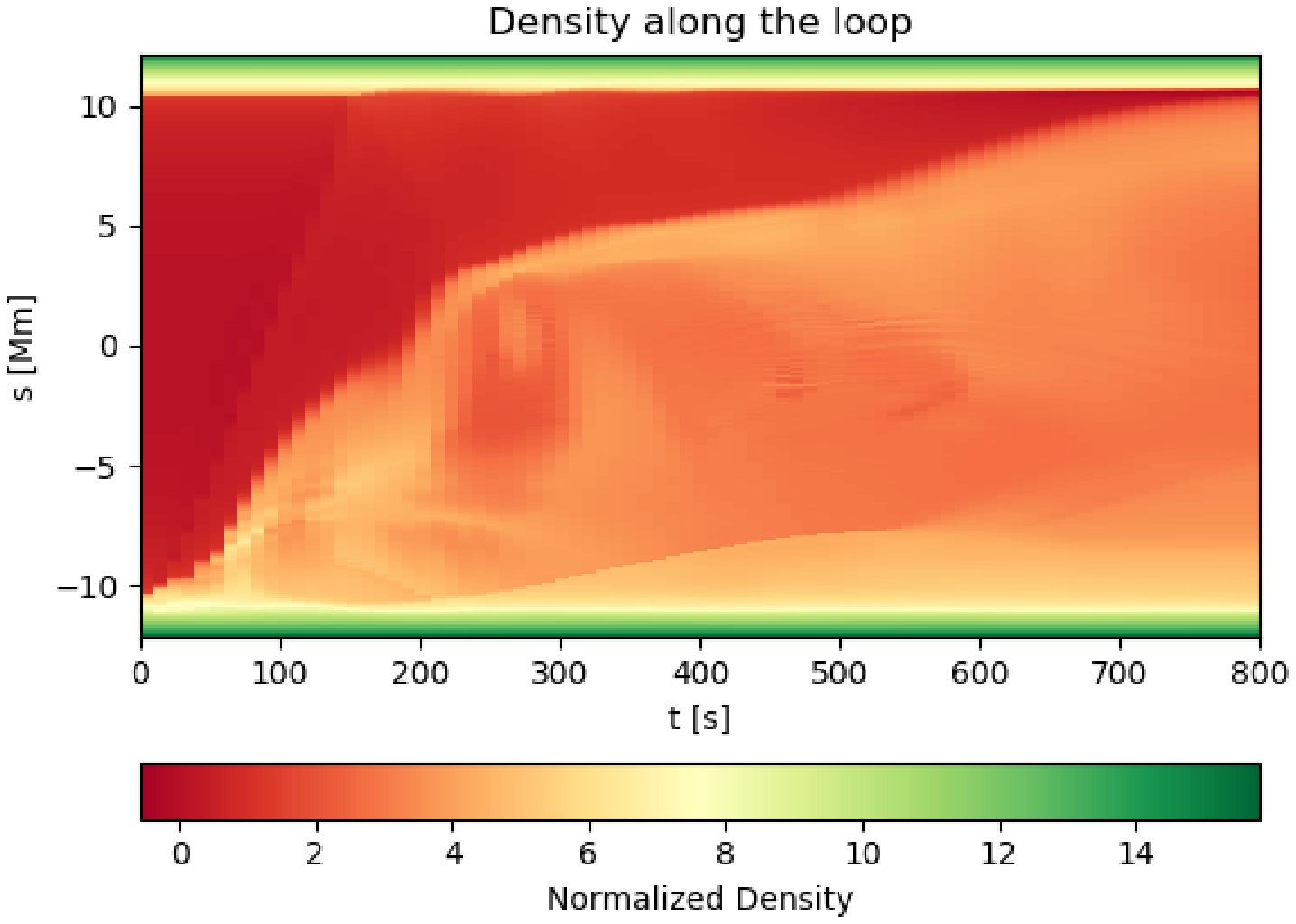}
\includegraphics[scale=0.4,angle=0,width=8cm,height=12cm,keepaspectratio]{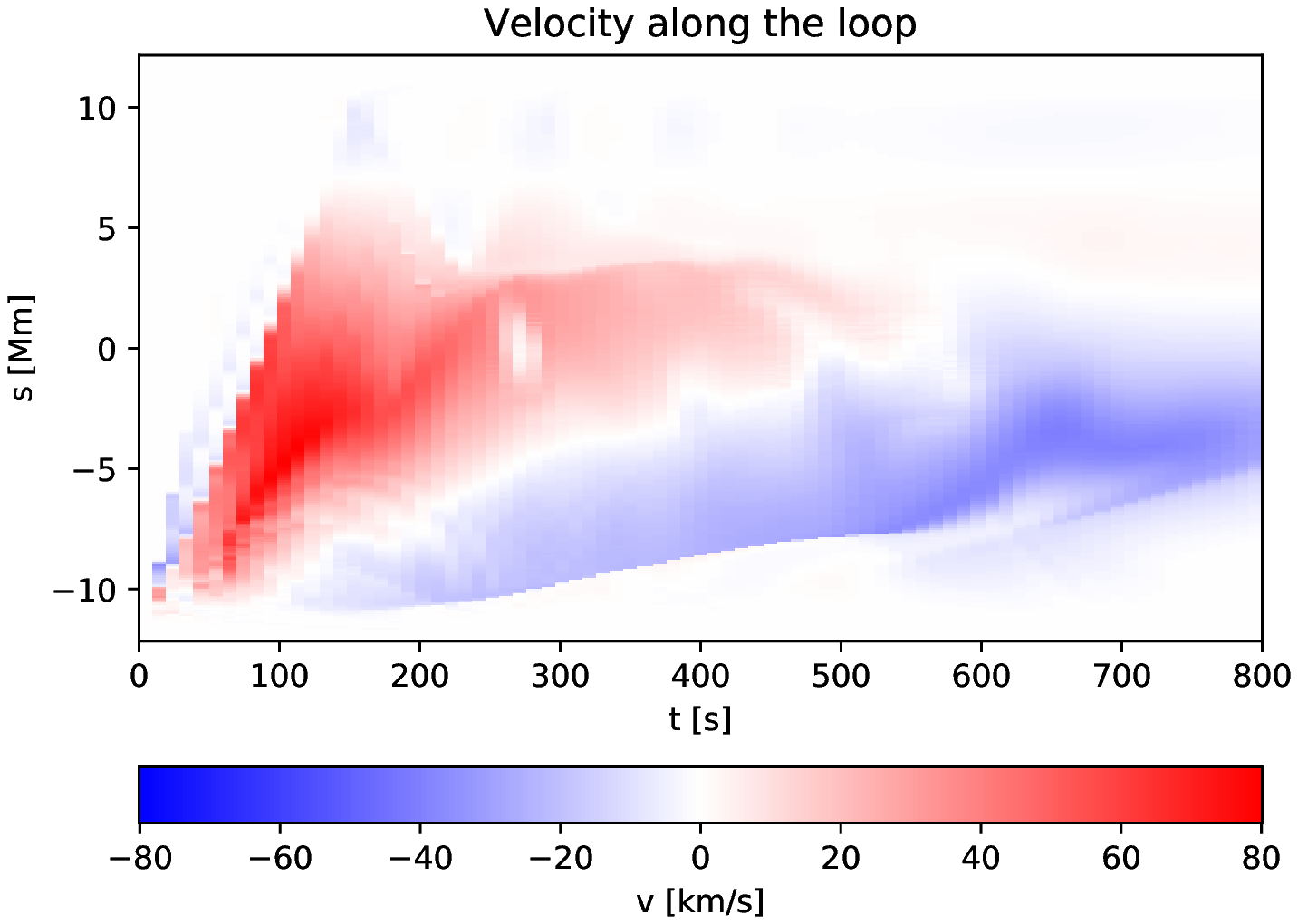}
}
\label{mass_evolve}
\caption{Temporal evolution of the mass density (left) and flows (right) along the chosen set of the curved field lines along which the cool loop system evolved.}
\end{figure*}
To visualize the fine structures in the model cool loop system, we have made A$_{ez}$ vs $\phi{_e}$ transformation of the field of view (x-y coordinates) of the simulation data and straightened the curved field lines along with the interpolated physical quantities as shown in Fig.~\ref{evolve_fine}. Eqs.~10-12 exhibit the choice of the magnetic fields in x-y space (cf., Figs.~6-7). The equipotential lines are perpendicular to the magnetic lines. $\phi{_e}$=constant represents the  equipotential line and A$_{ez}$=constant is the lines of magnetic force. Fig.~\ref{evolve_fine} displays the plasma flows from the left footpoint of the loop system to its right footpoint, and the system evolves subsequently. These density maps show straightened view of the simulated plasma flows in the cool loop. These simulated loops exhibit striking similarities with the observed loop system in Fig.~\ref{1}. The formation of fine structured flowing plasma near both footpoints is clearly visible in these maps which match with the observational data (Movie 1). This clearly depicts that the perturbed plasma moves from one footpoint and reaches to the another footpoint, and lots of fine structured mass motions are developed within this closed loop system especially near the footpoints (Fig.~8).  

Fig.~9 presents the temporal evolution of the normalized mass density (left-panel) and velocity (right-panel) along the path of flowing plasma in the cool loop system. The mass density is normalized by PLUTO unit density, i.e 10$^{-15}$ g cm$^{-3}$. 
The estimated velocity (V) field is the resultant of V$_{x}$ and V$_{y}$ components of the velocities in the 2-D simulation box. Along the loop system, after implementing the velocity pulse, there is an obvious motion of the cool plasma with certain velocities at different spatial points. Ahead of this mass motion, the magnetoacoustic shock has propagated and rebound from the other footpoint of the loop (cf., Fig.~8). We use  0.0 km s$^{-1}$ velocity as a reference velocity to estimate the Doppler velocity patterns. We have computed the Doppler velocity using data of signed total velocity estimated along the chosen flowing cool loop system, and we have plotted its temporal variation (Fig.~9).  The sign convention is that the velocity field which is directed from left footpoint upward will possess negative sign of the velocity (blue-shift), while the converse will represent the downward directed velocity vectors with the positive sign (red-shift). Up to $\approx$100 s, the velocity perturbations grow rapidly and form a magntoacoustic shock along the loop that is propagating ahead and rebound back rapidly from the another footpoint. Therefore, initially up to $\approx$100 s the entire loop is associated with the red-shifted velocity field (backward/oppositely directed), which basically represent that the loop thread is quickly filled with the rebound shock. At the time when shock is in the process of rebounding back from the right footpoint of the loop, the cool plasma follows the shock behind and evolves above the left footpoint due to the creation of the low pressure region. In this way, the flow has started that further forms the dense loop thread. In such a scenario, the upward flowing plasma also tries to counteract the rebound shock. Therefore, after $\approx$100 s the blue-shifted segment rises up along the loop length, while the red-shifted segment starts diminishing in the velocity map (Fig.~9b). As time progresses, the cool plasma flows in the loop, and it fills in the entire loop system. Due to the interaction of the flowing plasma from left footpoint to right, and their interaction with the counter-propagating shock signals, leading to lots of fine structured plasma segments following the magnetic field lines (Fig. 8).

\section{Discussion and Conclusions}

The solar chromosphere is a complex magneto-plasma system that offers the evolution of a variety of plasma dynamics (e.g., jets, mass motions, shocks) as well as magnetic waves.  It is separated by the inner corona through the particular discontinuities (e.g., mass, density, temperature, characteristic speeds, etc). Therefore, a lot of phenomena are seen below the inner corona in the form of their reflection, conversion, and dissipation. Flowing cool loop systems are always very important magnetic structures because of their capability to guide the transport of the plasma from the lower to the upper solar atmosphere. If such flowing tubes interact with the neighbourhood open magnetic domains, they can transfer the flowing mass to the open magnetic channels. Understanding the occurrence of the impulsive mass flow in such loops is always a difficult problem because more than one physical process may be responsible for it. Recently, \citet{2015ApJ...810...46H} have observed the formation of the flowing cool loop system and enhanced line-profiles at their footpoints. The effective velocity was observed to be as high as 200 km s$^{-1}$ over there. We have taken this observational aspect to initiate our model, where cool loop system is evolved efficiently almost similar the observed one by the implementation of similar velocity perturbations above one footpoint. The unusually broadened  line profiles at the footpoint of the loop system indicates the impulsive energy releases. However, in the present observational base-line, we only invoke the velocity responses of the EEs in the chromosphere/TR
that essentially form a flowing cool loop system.

There might be other possibilities for such flows which have been discussed in earlier studies (see e.g., \citealt{2014SoPh..289.2971P,2019ApJ...887...56T}). \citealt{2014SoPh..289.2971P} have shown that mixed polarity flux region, where the loops are anchored, might give flux cancellation signatures exhibiting plasma driven by the magnetic reconnection.
There are other possibilities also, e.g., sudden release of magnetic tension by ambipolar diffusion which can drive spicule like outflows at the loop footpoint causing the bulk flows (\citealt{2017Sci...356.1269M}). Siphon flow can simply drive outflow from one end of a loop if the field is significantly lower at that end causing a larger gas pressure there than the other end (where the field would be stronger resulting into a lower gas pressure at the photospheric height) as found in a loop by \citet{2015ApJ...810...46H}. \citet{2019ApJ...874...56R} have also discussed the plasma driven by impulsive heating targeting the same active region.
 
However, the present modeling work highlights the formation of the fine-structured cool loop system due to multiple episodic velocity enhancements, which may be most likely associated with the responses of the explosive events (EEs) in the upper chromosphere/TR as seen in the observations. These velocity enhancements/perturbations provide significant amounts of kinetic energy and momentum to the plasma. Since the complex interaction of the flowing plasma and rebound magnetoacoustic shock takes place in the curved magnetic fields, the fine structure of the flowing plasma is evident in the cool loop system. Our finding depicts the numerical modelling of the plasma driven by the localised velocity enhancements/perturbations at one of the footpoint in the cool loop system most likely due to the response of EEs. Such physical scenario has been discussed by \citet{2019ApJ...874...56R} as a strong possibility for such flows for one of the dataset targeting the same active region. We compare the spatio-temporal evolution of the flowing cool loop system in the framework of our model with the observed one, and conclude that their formation is mostly associated with the transient energy release at their footpoints in the chromosphere/TR. Therefore, such velocity responses, which may be most likely associated with the EEs, are found to be the main candidates for the mass evolution and energetics of the flowing cool loop systems in the lower solar atmosphere. Future discussions are open to numerical modeling of the flowing cool loops by implementing other physical mechanisms as already mentioned before. It is quite likely that more than one mechanism (such as siphon flow in addition to the impulsive heating) may be at work. Such an extensive investigation is underway.

%
%
\section{Acknowledgements}
We thank the reviewer for his/her constructive comments which improved our manuscript considerably. IRIS is a NASA small explorer mission developed and operated by LMSAL with mission operations executed at NASA Ames Research center and major contributions to downlink communications funded by ESA and the Norwegian Space Centre. The IRIS data are publicly available from the Lockheed Martin Solar and Astrophysics Laboratory (LMSAL) website (http://iris.lmsal.com/). The open source SolarSoft code package (http://www.lmsal.com/solarsoft/) is used for the initial data processing. S.K.T. gratefully acknowledges support by NASA contracts NNG09FA40C (IRIS), and NNM07AA01C (Hinode). AKS and YKR thank Dr. Pradeep Kayshap for some of his suggestions on the spectroscopic analyses. AKS and MM acknowledge the support of UKIERI grant for the support of their research. Armagh Observatory and Planetarium is core funded by the Northern Ireland Executive though the Department of Communities. Authors
also acknowledge the use of PLUTO MHD code in the present scientific work.

%
%
%

\end{document}